\documentclass[%
 reprint,
superscriptaddress,
 amsmath,amssymb,
 prb,
floatfix,
]{revtex4-1}

\usepackage{color}

\usepackage{graphicx}
\usepackage{dcolumn}
\usepackage{bm}
\usepackage{mhchem}
\usepackage{nicefrac}

\newcommand{\la}{LaFe$_2$As$_2$}
\newcommand{\ba}{BaFe$_2$As$_2$}
\newcommand{\ca}{CaFe$_2$As$_2$}

\begin{document}

\title{Novel Fe-based superconductor {\la} in comparison with traditional pnictides}

\author{I. I. Mazin}
\affiliation{Code 6393, Naval Research Laboratory, Washington, DC 20375, USA}

\author{Makoto Shimizu}
\affiliation{Department of Physics, Okayama University, Okayama 700-8530, Japan}

\author{Nayuta Takemori}
\affiliation{Research Institute for Interdisciplinary Science, Okayama University, Okayama 700-8530, Japan}

\author{Harald O. Jeschke}
\affiliation{Research Institute for Interdisciplinary Science, Okayama University, Okayama 700-8530, Japan}

\date{\today}

\begin{abstract}
The recently discovered Fe-based superconductor (FeBS) {\la} seems to 
break away from an established pattern that doping FeBS beyond $0.2 e$/Fe 
destroys superconductivity. {\la} has an apparent doping of $0.5 e$, 
yet superconducts at 12.1~K. Its Fermi surface bears no visual 
resemblance with the canonical FeBS Fermiology. It also exhibits two 
phases, none magnetic and only one superconducting. We show that the 
difference between them has nonetheless magnetic origin, the one 
featuring disordered moments, and the other locally nonmagnetic. We find 
that La there assumes an unusual valence of $+2.6$ to $+2.7$, so that the 
effective doping is reduced to $0.30$-$0.35 e$. A closer look reveals the 
same key elements: hole Fermi surfaces near $\Gamma$-$Z$ and electron ones near 
the $X$-$P$ lines, with the corresponding peak in susceptibility, and a 
strong tendency to stripe magnetism. The physics of {\la} is thus 
more similar to the FeBS paradigm than hitherto appreciated.
\end{abstract}


\maketitle

For more than a decade after the discovery of Fe-based superconducting pnictides\cite{Kamihara2008} it seemed that superconductivity existed in a relatively narrow range of dopings away from the nominal Fe$^{2+}$ valency, between $-0.2e$ and $0.15e$, and disappeared or was rapidly suppressed after that~\cite{Kordyuk2012}. It was rationalized in terms of the Cooper pair scattering between the hole pockets of the Fermi surface near the zone center and electron pockets near its corner\cite{Hirschfeld2011}. One of the few examples of strongly overdoped (up to Fe$^{1.5+}$ valency) pnictides was provided by Hosono's group~\cite{Hanna2011,Iimura2012,Hiraishi2014}, namely the 1111 material LaFeOAs, with up to 50\% of O$^{2-}$ replaced by H$^-$. Intriguingly, they observed two superconducting domes, possibly with different pairing symmetries and/or mechanisms. Unfortunately, further study of this material, both theoretical and experimental, has been hindered by the volatility of hydrogen and natural disorder.

Recently, another compound with formally Fe$^{1.5+}$ has been synthesized~\cite{Iyo2019}, {\la} (La122), stoichiometric and isostructural with the arguably best studied iron pnictide, {\ba}. It was found experimentally that the material can exist in two distinctly different crystallographic phases: as-prepared samples exhibit a 6.5\% shorter crystallographic $c$-parameter and a 1.6\% shorter $a$-parameter than the same sample, annealed. In  Ref.~\onlinecite{Iyo2019} they were named ``collapsed-tetragonal'' (CT) and ``uncollapsed-tetragonal'' (UT).  The latter, but not the former, was exhibiting superconductivity at 12.1~K.
A similar $c$ parameter collapse had been observed in the {\ca} (Ca122) compound, where it is triggered by pressure and is accompanied by the formation of As-As dimers~\cite{Johrendt1997,Tomic2012}, and by reduction of $c$ and small increases of $a$ and $b$. The formation of the As-As bond was initially considered to be the driving force for the collapse~\cite{Yildirim2009}, but a later investigation of Ca$_{1-x}$Sr$_x$Fe$_2$As$_2$ suggested~\cite{Zhao2018} that what drives the collapse is the loss of magnetism, while the As dimerization is a byproduct. At first glance, neither idea is applicable to La122, because the As-As bond in the CT phase is considerably longer than in Ca122 (3.18~{\AA} vs. 2.84~{\AA}), and the experiment does not show any ordered magnetism in the UT phase.
Furthermore, the Fermi surface calculated in Ref.~\onlinecite{Iyo2019}, on the first glance, bears no resemblance with that of the traditional iron pnictides; especially the ubiquitous hole pockets near the zone center seem to be absent.

In this contribution, we will address structural, magnetic, and electronic properties of La122, and will show that the UT phase, as opposed to the CT one, carries a strong short-range magnetism of the stripe type driven by the next-nearest-neighbor exchange, with, however, different subdominant interactions. 
The structural changes are driven by the magnetic collapse, as in Ca122. The orbitals relevant for the low-energy physics are not the usual $d_{xz}$ and $d_{yz}$ (these are nearly completely filled) but $d_{xy}$ and $d_{z^2}$ instead. The former forms a quasi-2D cylinder at the zone center, which, contrary to the initial assertion, is quite similar to the hole pocket in traditional iron pnictides. This fact was overlooked because a very 3D $d_{z^2}$ band forms a Fermi surface sheet that crosses the $d_{xy}$ cylinder and hybridizes with it, hiding it from view if every sheet is plotted separately. Most importantly, La in this compound assumes a non-integer valence closer to 2.7+ than to 3+, corresponding to doping of $\sim 0.35e$, rather than $0.5e$; 
it is thus overdoped, but not dramatically. The overextended electron pockets do not exclude the usual spin-fluctuation driven mechanism with an overall $s$ symmetry. However, in such a scenario the order parameter will nearly necessarily be nodal.

\begin{table}
  \begin{ruledtabular}
    \begin{tabular}{llll}
      structure & $a$~({\AA}) & $c$~({\AA}) & $z_{\rm As}$\\\hline
UT, exp. &  3.9376 & 11.7317 & 0.3657\\
stripe, theor.&  3.9281 & 11.7960 & 0.3620 \\\hline
CT, exp. & 4.0035 & 11.0144 & 0.3589\\
nonmagnetic, theor. & 3.9904 & 10.9535 & 0.3573
\end{tabular}
  \end{ruledtabular}
  \caption{Structural parameters of {\la} as determined experimentally (Ref.~\onlinecite{Iyo2019}) and obtained by full structural relaxation.}\label{tab:structure}
\end{table}

{\it Methods.-} We use density functional theory (DFT) with a projector augmented wave basis as implemented in VASP~\cite{Kresse1993} for structure prediction. All calculations employ a generalized gradient approximation (GGA) exchange correlation functional~\cite{Perdew1996}. We use the all electron full potential local orbital (FPLO) basis~\cite{Koepernik1999} for electronic structure analysis~\cite{Shimizu2018} and energy mapping of magnetic states~\cite{Glasbrenner2015,Guterding2017}. Based on a tight-binding model from projective Wannier functions~\cite{Eschrig2009}, we determine a non-interacting susceptibility~\cite{Graser2009,Guterding2015a}.

First, we address the structural properties, using density functional theory (DFT) as a tool. In a similar compound, LaFe$_2$P$_2$, DFT was shown to reproduce the experiment in much detail, arguably better that in Fe$^{2+}$ pnictides~\cite{Muranaka2009,Blackburn2014}. We have performed the full optimization of the crystal structure, using two approaches: one spin-unrestricted, and the other enforcing a nonmagnetic Fe state. In the former we have found the same stripe-order ground state as in {\ba}, with sizable magnetic moments of 1.8~$\mu_{\rm B}$ inside the Fe PAW sphere, slightly smaller than in {\ba}. The structural parameters, listed in Table~\ref{tab:structure}, were extremely close to those experimentally determined for the UT phase. In contrast, the nonmagnetic calculations converged to a structure nearly identical with the experimental CT structure. It had been already well established (see, for instance, Ref.~\onlinecite{Mazin2008a}) that in order to reproduce the crystal structure of paramagnetic iron pnictides one needs to account for the fluctuating local magnetic moments by allowing an appropriate magnetic order. Otherwise, the Fe$^{2+}$ ionic radius is too small, the Fe-As bond too short, and $c/a$ too small as well. The only material where nonmagnetic calculations generate a correct structure is the CT phase of Ca122.

In pure Ca122 one cannot disentangle  the effect of magnetism and the effect of As-As bonding; it had been generally believed that both contribute to the collapse,and one piece of evidence of the contrary\cite{Zhao2018} has not been universally accepted. Our result clearly shows magnetism in the driver seat. Experimentally and theoretically, As-As bonding is much weaker in CT La122 than in CT Ca122, yet their structural characteristics are amazingly similar.

\begin{figure}
\includegraphics[width=0.48\textwidth]{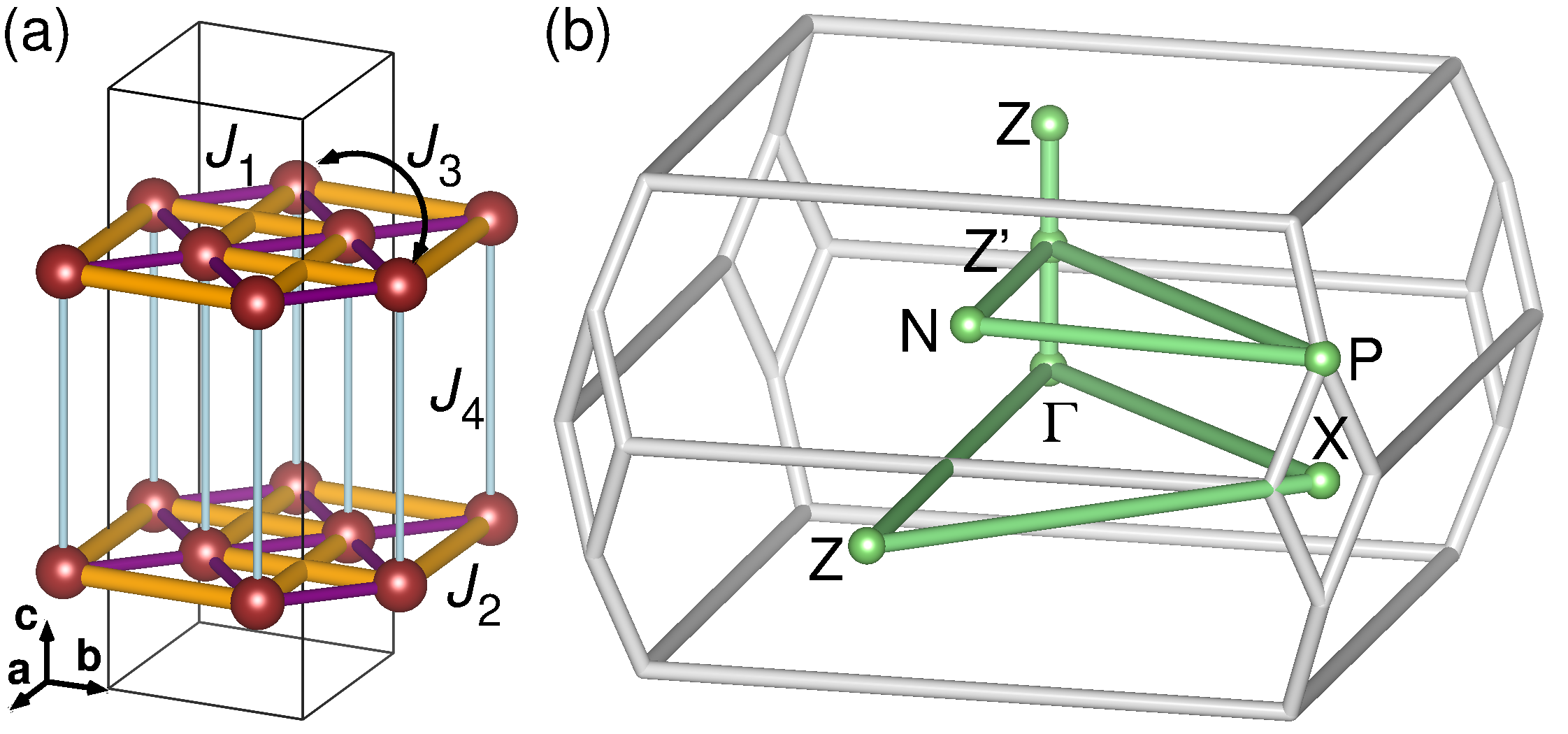}
\caption{(a) Relevant exchange paths in {\la}. (b) Brillouin zone of the $I4/mmm$ space group of {\la}.}\label{fig:bz}
\end{figure}

Having established this fact, we have analyzed the magnetic interaction by fitting the calculated total energies in the UT phase onto the four nearest neighbor couplings of the Heisenberg Hamiltonian (the paths are shown in Fig.~\ref{fig:bz}~(a)). The results are presented in Table~\ref{tab:exchange}. Similar to other pnictides, the next nearest neighbor interaction $J_2$ (the one supporting the $s_\pm$ pairing~\cite{Seo2008}) dominates. Unlike those other materials, $J_1$ is not only smaller than $2J_2$, assuring the stripe order, but is negligible. Furthermore, the third neighbor interaction, responsible for the double-stripe order in FeTe, is also quite sizable in La122, but is ferro- rather than antiferromagnetic. It remains to be seen whether such an unusual behavior would follow from the standard low-energy itinerant model~\cite{Fernandes2017} and what particular effect it may have on the pairing symmetry, but it is likely to modify the pairing interaction in a substantial and interesting way.

\begin{table}
  \begin{ruledtabular}
    \begin{tabular}{crrrr}
material& $J_1$~(K) & $J_2$~(K) & $J_3$~(K)  & $J_4$~(K)\\\hline         
{\la}   & 25(13) & 306(7) & -108(13) & 63(19) \\
{\ba}   & 75(16) & 400(23) &  -65(40) &  151(8) 
\end{tabular}
  \end{ruledtabular}
  \caption{Exchange couplings of {\la} and {\ba} calculated by mapping GGA total energies of eleven spin configurations to a Heisenberg Hamiltonian. The exchange paths are visualized in Fig.~\ref{fig:bz}~(a).}\label{tab:exchange}
\end{table}

\begin{figure}
\includegraphics[width=0.48\textwidth]{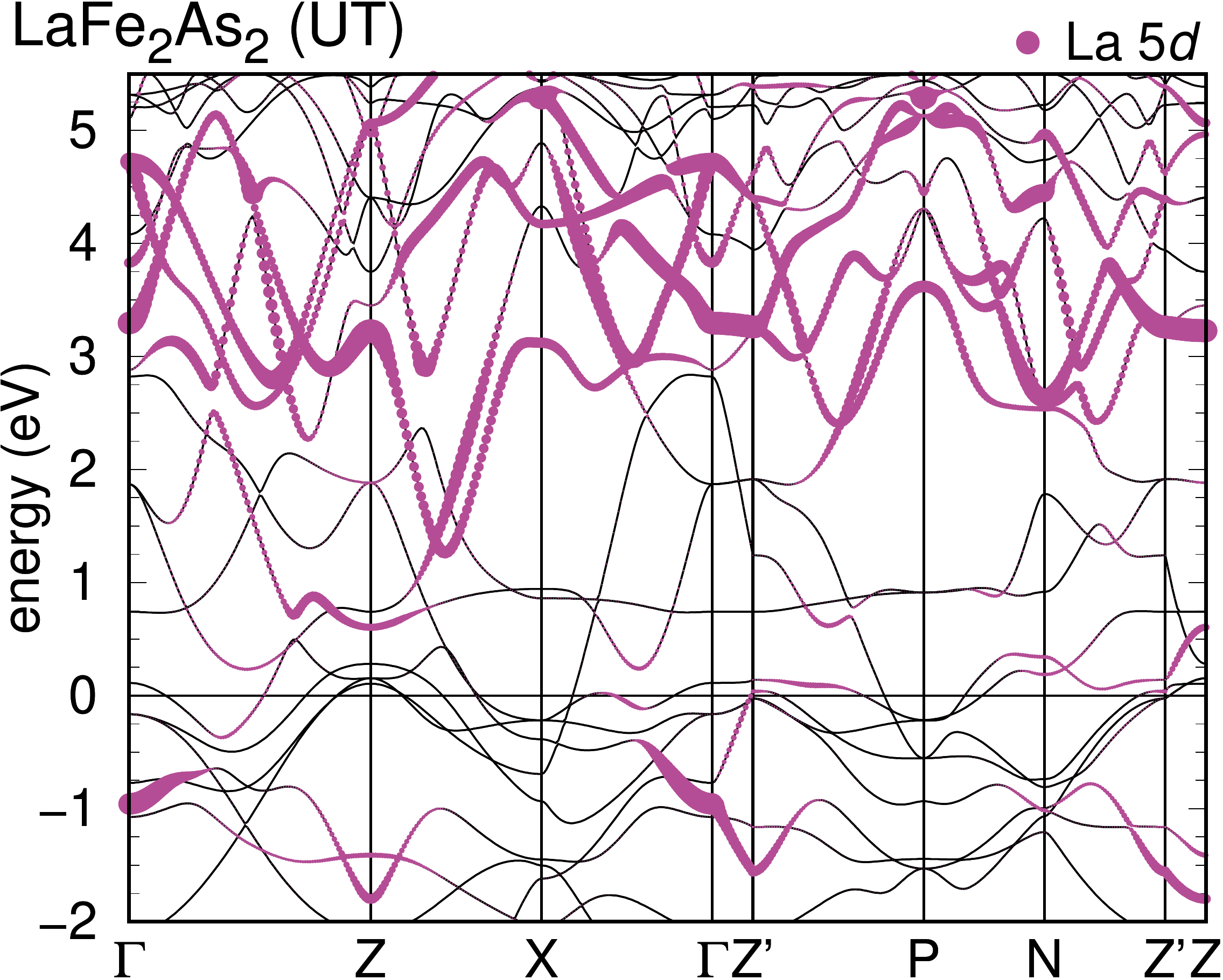}
\caption{Band structure of UT {\la} along the path shown in Figure~\ref{fig:bz}~(b) with total La $5d$ weights marked. The occupied La weight is mostly of $5d_{xy}$ character.}\label{fig:Lastates}
\end{figure}

\begin{figure*}
\includegraphics[width=\textwidth]{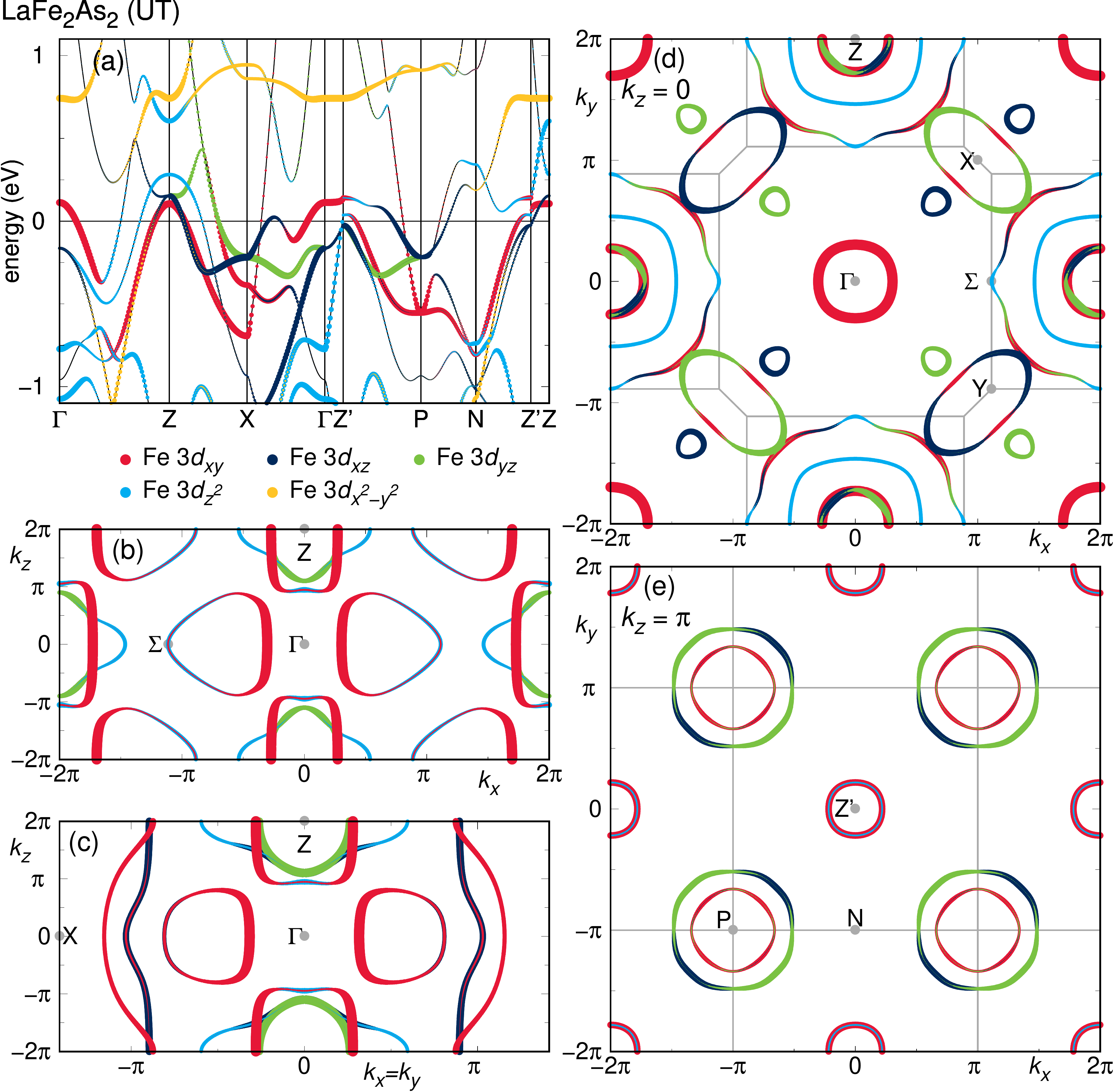}
\caption{Band structure and Fermi surfaces of UT {\la} calculated within GGA. The high symmetry points are standard body centered tetragonal points as listed in Ref.~\protect\onlinecite{Setyawan2010}, except for $Z'$ which is half way between $\Gamma$ and $Z$ (see the path marked in Figure~\ref{fig:bz}~(b)).}\label{fig:UTbsfs}
\end{figure*}

To this end, let us now present a detailed description of the electronic structure in the paramagnetic UT phase. As usual, the calculations are performed in the nonmagnetic case, deemed to be a good approximation to the paramagnetic state.

 Fig.~\ref{fig:Lastates} shows the calculated band structure with the La $5d$ character highlighted, and Fig.~\ref{fig:UTbsfs} shows the bands and Fermi surface cuts with Fe $3d$ characters.

First and foremost, we see that one of the two La-$5d$ $e_g$ bands, $5d_{xy}$, which is strongly hybridized with As $4p$, has a huge dispersion, from its bottom at the $\Gamma$ point at $-1.5$ eV, to the top at $X$ at $+5$ eV (Fig.~\ref{fig:Lastates}). Because of that, it becomes partially occupied, and, even though there are no pure La bands at the Fermi level, it absorbs some number of electrons, noticeably reducing the effective doping. Because of strong hybridization between this band and Fe orbitals,there is no rigorous way to assess this reduction. We have used two methods, one of which is supposed to give a lower bound, and the other the upper bound. In both cases we started from a tight-binding fit with all orbitals but Fe $3d$ and La $5d$ integrated out. The fit is not perfect around the $X$ point, but pretty good closer to $\Gamma$ and $Z'$. Then we zero non-diagonal elements between La and Fe and calculate the number of electrons in the (now pure) La $5d$ band. This gives us $0.22 e$ per La. Next, we take the original fit, with Fe-La hybridization, and integrate the La $5d_{xy}$ partial density of states. This gives $0.47 e$ per La. We consider these two numbers to be the lower and the upper bound, with $0.3$-$0.4 e$ the most likely number. Note that this corresponds to the actual Fe doping of $0.30$-$0.35e$/Fe, which is, of course, past optimal, but not nearly as heavy as $0.5e$. 

Next, let us take a closer look at the Fe bands near the $\Gamma$ point. We see that the $d_{xz}/d_{yz}$ bands that play the leading role in the Fe$^{2+}$ pnictides are now nearly entirely below the Fermi level in the vicinity of the $\Gamma$ point (they do form a tiny 3D hole pocket around $Z$, which is basically irrelevant). This does not mean that the essential hole cylinder around $\Gamma$-$Z$ is gone; the $d_{xy}$ band, which appears in many, albeit not all Fe$^{2+}$ pnictides, is well visible near $\Gamma$, and the two cuts showing the vertical $\Gamma$-$Z$ direction (Figures~\ref{fig:UTbsfs}~(b) and (c)) show it to be nearly 2D. This band is a perfect candidate for the standard $s_\pm$ scenario. The reason that it was missed in Ref.~\onlinecite{Iyo2019} is that at $k_z\sim (\Gamma$-$Z)/2$  it is cut across by another, complicated Fermi surface, formed mostly by the $d_{z^2}$ orbital, which gets gradually mixed with other orbitals as $k_x,k_y$ increase.
Note that this orbital extends along $z$ and is thus more dispersive along $k_z$ than in the $k_x$-$k_y$ plane. The corresponding Fermi surface is therefore very three-dimensional. The $d_{z^2}$ sheet hybridizes with the $d_{xy}$ cylinder, creating a visually complicated topology, which, however, can be readily traced down to these two elements. 

Having established the existence of a sizable quasi-2D hole pocket near the $\Gamma$-$Z$ line, let us see whether we can reveal electron states sufficiently close to the $X$ point to recover the standard pairing scenario~\cite{Mazin2008} (note that because of the 3D character of the electronic structure we are using the standard notations for the body-centered tetragonal symmetry; point $X$ in this notation corresponds to the $M$ point in the often-used 2D nomenclature). 

Again, let us begin with the $d_{xz}/d_{yz}$ bands. At $X$ they sit at respectable 220~meV below the Fermi level, and disperse upward pretty much in the same way as they do in other iron pnictides. They strive to form a large Fermi surface cylinder ($k_{\rm F}\sim \pi/2a$), but this is interrupted by hybridization with other bands, $d_{xy}$ and $d_{z^2}$. Right in the middle between $\Gamma$ and $Z$, this hybridization is absent and the Fermi surface cut at this $k_z$ looks amazingly similar to the Fermi surface topology in typical iron pnictide superconductors (Figure~\ref{fig:UTbsfs}~(e)). Given that the usual $d_{xz}/d_{yz}$  hole pockets are absent, these states are unlikely to play a leading role in superconductivity. 

Interestingly, the $d_{xy}$ band also shows up near $X$. While at $X$ it is located deep below the Fermi level ($\sim 0.8$ eV), it disperses upward extremely rapidly, and along $\Gamma$-$X$, where it cannot hybridize with the $d_{xz}/d_{yz}$ bands by symmetry, it crosses the Fermi level already at 0.2 of the distance between $\Gamma$ and $X$.  Again, this simple Fermiology is disrupted by hybridization with other states, except along $\Gamma$-$Z$. However, it does not nullify the fact that there are plenty of $d_{xy}$ states in the electronic pockets around the $X$-$P$-$X$ vertical line, which share the character with zone-center hole states, and can lead to the same $s_\pm$ superconductivity as in Fe$^{2+}$ pnictides, despite a different doping level and visually extremely different Fermi surface.

\begin{figure}
\includegraphics[width=0.48\textwidth]{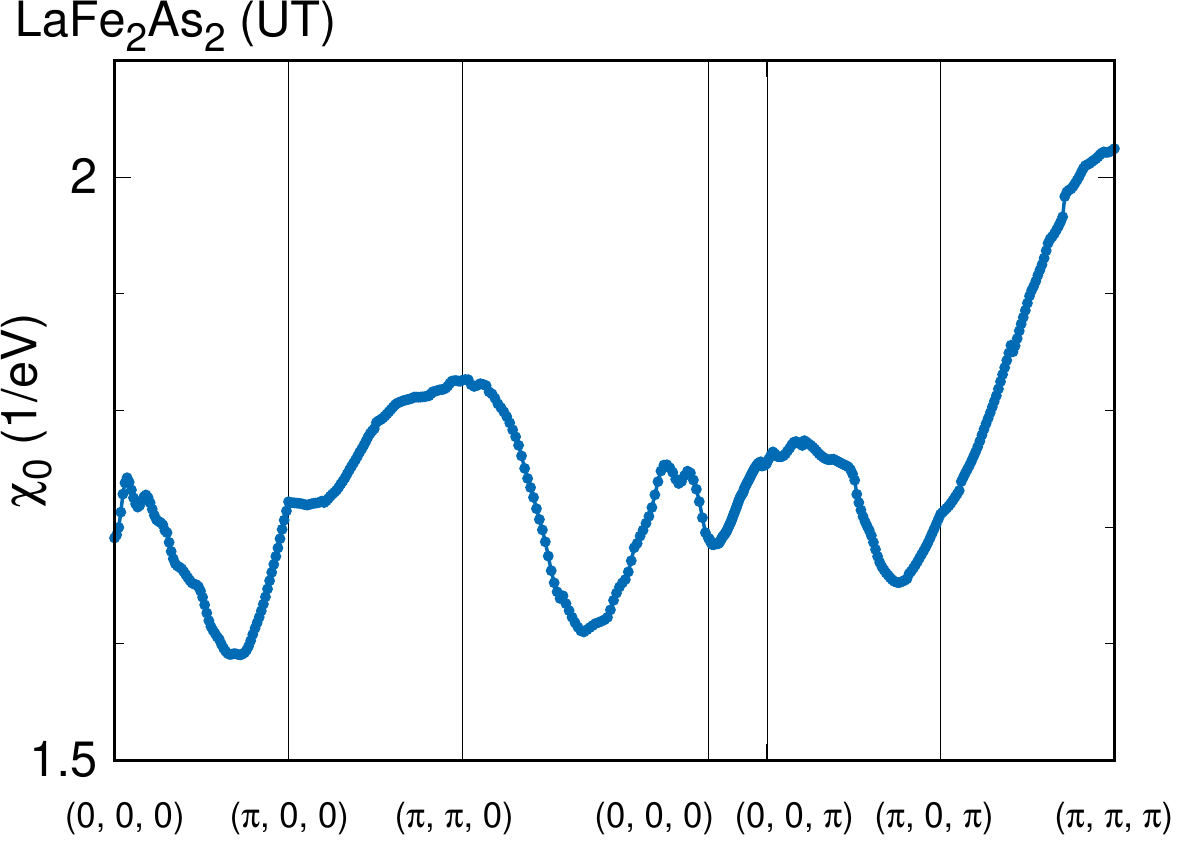}
\caption{Noninteracting susceptibility of UT {\la}.}\label{fig:susc}
\end{figure}

This description is very reminiscent of the well-known scenario for superconductivity in optimally doped iron pnictides, except that instead of the $d_{xz}/d_{yz}$ band showing up in both hole and electron pockets, we have $d_{xy}$. With this in mind, we have calculated the noninteracting susceptibility, including the orbital-defined matrix elements. It is displayed in Figure~\ref{fig:susc}. As expected, there is a large peak near ${\bf Q}=(\pi/a,\pi/a,q_z)$, both at $q_z=0$ and even stronger at $q_z=\pi/c$. Together with the fact that the calculated mean-field ground state is strongly stripe-type antiferromagnetic, it convincingly suggests that the dominant spin fluctuation has a $(\pi/a,\pi/a)$ wave vector. Note that the calculations provide a sizable antiferromagnetic interlayer coupling, conforming to a global susceptibility maximum at ${\bf Q}=(\pi/a,\pi/a,\pi/c)$.

To summarize, we find that while {\la} is unquestionably a unique, unusual and highly interesting material, many of its apparent mysteries may have simple resolutions. First, the strange coexistence of two structurally different phases finds explanation in different local magnetic states of Fe ions -- despite the absence of a long range order in either. Second, the real doping level of Fe bands is considerably smaller than the one derived from a purely ionic picture. Third, the Fermi surface of {\la} is indeed very complex, but this complexity hides the same basic motif as in traditional iron based superconductors: a hole pocket near $\Gamma$ and electron pockets near $X$. Fourth, spin fluctuations are also peaked near ${\bf Q}=(\pi/a,\pi/a,q_z)$, potentially providing the necessary superconducting ``glue''.

\acknowledgments
I.I.M. acknowledges support by ONR through the NRL basic research 
	program and by the Research Institute for Interdisciplinary Science, Okayama University visiting scientist program.

\end{document}